\documentclass[aps,prl,twocolumn,showpacs]{revtex4}
\usepackage{txfonts}
\usepackage{graphicx}

\begin{document}

\title{Nodeless Superconductivity in the Noncentrosymmetric Superconductor $Mg_{10}Ir_{19}B_{16}$}

\author{Gang Mu, Yue Wang, Lei Shan and Hai-Hu Wen}\email{hhwen@aphy.iphy.ac.cn }

\affiliation{National Laboratory for Superconductivity, Institute of
Physics and Beijing National Laboratory for Condensed Matter
Physics, Chinese Academy of Sciences, P.O. Box 603, Beijing 100080,
People's Republic of China}

\begin{abstract}
We measured the resistivity, diamagnetization and low temperature
specific heat of the newly discovered noncentrosymmetric
superconductor $Mg_{10}Ir_{19}B_{16}$. It is found that the
superconducting gap has an s-wave symmetry with a value of about
$\Delta_0 \approx$ 0.6 meV, and the ratio
$\Delta_0/k_BT_c\approx1.86$ indicates a weak coupling for the
superconductivity. The correlations among the normal state
Sommerfeld constant $\gamma_n$, the slope $-d\mu_0H_{c2}(T)/dT$ near
$T_c$ and the condensation energy $E_c$ are all consistent with the
weak coupling picture. The separated phonon contribution from the
specific heat shows that the conduction electrons of the $Ir$ atoms
interact most strongly with the vibrations of themselves, instead of
with that of the light element boron.
\end{abstract}
 \maketitle

The study on superconductivity in noncentrosymmetric materials has
attracted growing efforts in recent
years\cite{GorkovPRL2001,FrigeriPRL2004,EdelsteinJETP1989,LevitovJETP1985,SamokhinPRB2004}.
For most superconductors, the atomic lattice has a centrosymmetry,
therefore the system is inversion symmetric. The orbital part of the
superconducting order parameter has a subgroup which is confined by
the general group of the atomic lattice. Due to the Pauli's
exclusion rule and the parity conservation, the Cooper pair with
orbital even parity should have anti-parallel spin state, namely
spin singlet, while those having orbital odd parity should have
parallel spin state, i.e., spin triplet. If a system lacks the
centrosymmetry, the triplet pairing may be instable leading to a
mixture of singlet and triplet pairing. Theoretically novel features
are anticipated in the noncentrosymmetric
system\cite{HayashiPhysicaC2006}. A nodal gap structure has been
observed in $Li_2Pt_3B$ showing the possibility of triplet pairing,
while due to weaker spin-orbital
coupling\cite{YuanHQPRL,ZhengGQPRL}, the nodal gap has not been
observed in a material $Li_2Pd_3B$ with similar structure. It is
thus highly desired to investigate the paring symmetry in more
materials with noncentrosymmetric structure.

The newly discovered superconductor $Mg_{10}Ir_{19}B_{16}$
(hereafter abbreviated as $MgIrB$ )\cite{Klimczuk} with
superconducting transition temperature $T_c \approx 5$ K is one of
the rare materials which have the noncentrosymmetry. This material
has a space group of $I-43m$ with large and complex unit cells each
has about 45 atoms. To some extent it resembles the system
$Li_2(Pt,Pd)_3B$ since it has alkaline metals (Li, Mg), heavy
transition elements (Pd, Pt, Ir) and the light element boron.
Theoretically it was shown that the major quasiparticle density of
states (DOS) derives from the d-orbital of the heavy transition
elements. In this paper we present a detailed investigation and
analysis on the superconducting properties, such as the energy gap,
pairing symmetry, electron-phonon coupling strength and condensation
energy etc. in $MgIrB$. Our results suggest that the
superconductivity in this system is of BCS type with an s-wave gap
symmetry and a weak coupling strength.

The samples were prepared in two steps starting from pure elements
of Mg (98.5\%), Ir (99.95\%) and B (99.999\%) using a standard
method of solid state reaction. Appropriate mixtures of these
starting materials were pressed into pellets, wrapped in Ta foil,
and sealed in a quartz tube with an atmosphere of 95\% Ar/5\% $H_2$.
The materials were then heated at 600 $^\circ$ C and 900 $^\circ$ C
for 40 min. and 80 min., respectively. After cooling down to room
temperature, the samples were reground and mixed with another 20\%
of Mg, then they were pressed into pellets and sealed in a quartz
tube with the same atmosphere as used in the first step. In this
process the sample was heated up to 900 $^\circ$ C directly and
maintained for 80 min. The synthesizing process here is similar to
the previous work reported by the Princeton group\cite{Klimczuk} but
still with some differences. For example, we used Mg powder instead
of flakes to make the mixture more homogeneous. In addition the
pressure in the sealed quartz tube may rise to nearly 4 atm at 900
$^\circ$ C, which may considerably reduce the volatilization of Mg
during the synthesis. The resistivity and the AC susceptibility were
measured based on an Oxford cryogenic system (Maglab-Exa-12). The
specific heat was measured on the Quantum Design instrument PPMS
with temperature down to 1.8 K and the PPMS based dilution
refrigerator (DR) down to 150 mK. The temperatures of both systems
have been well calibrated showing consistency with an error below
2\% in the temperature range from 1.8 K to 10 K.

\begin{figure}
\includegraphics[width=8cm]{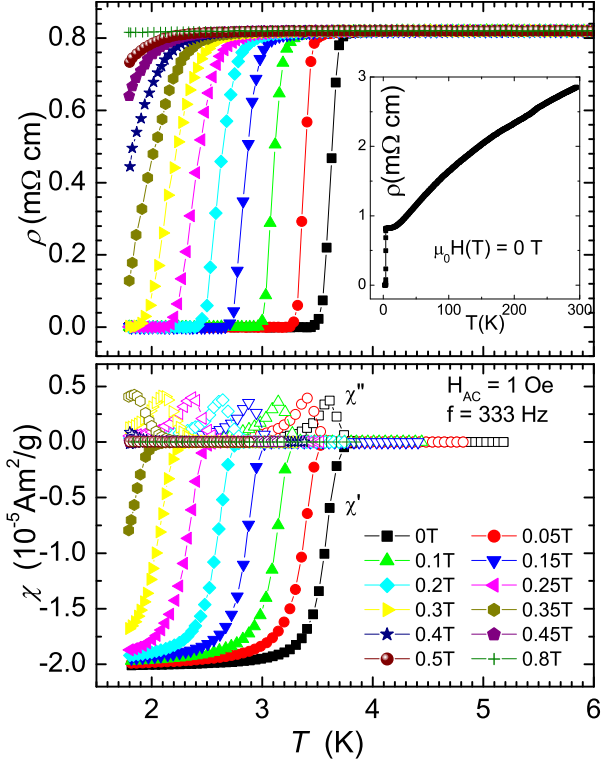}
\caption {(color online) Temperature dependence of resistivity (top)
and magnetic susceptibility ($\chi$" and $\chi$') (bottom) under
different DC magnetic fields. It is clear that the DC magnetic field
makes the transition shift parallel to low temperatures, manifesting
a field induced pair breaking effect. The inset in top panel shows
the resistive transition in a wide temperature regime at $\mu_0H$ =
0T.} \label{fig1}
\end{figure}

The x-ray diffraction (XRD) patterns taken on these samples show a
single phase with very small amount of impurity which is comparable
to that reported previously\cite{Klimczuk}. After the first round of
synthesizing, the superconducting transition inspected by the AC
susceptibility occurs at about 5 K with a relatively wide
transition. However, after the second step, the transition moves to
about 3.7 K with a sharper transition width. This indicates a
sensitive dependence of $T_c$ on the Mg content. In Fig. 1(a) we
show the temperature dependence of resistivity under different
magnetic fields. It is seen that the transition width ($1\% - 99\%
\rho_n$ ) is about 0.2 K. By applying a magnetic field the
transition shifts to lower temperatures quickly with a rather low
slope $-d \mu_0H_{c2}(T)/dT|_{T_c} \approx$ 0.3 $T/ K$. Using the
Werthamer-Helfand-Hohenberg relation\cite{WHH} $\mu_0H_{c2}=-0.69d
\mu_0H_{c2}(T)/dT|_{T_c} T_c$, we get the upper critical field
$\mu_0H_{c2}$ = 0.77 T. The AC susceptibility is shown in Fig.1 (b)
revealing a similar behavior as the resistive transition. It is
interesting to note that the value of the slope $d\mu_0H_{c2}(T)/dT$
found here is lower than that in the earlier report\cite{Klimczuk}
showing the tunability of superconducting properties in this system.

Shown in Fig.2 are the raw data of the specific heat. The open
squares represent the data taken with PPMS, while all filled symbols
show that taken with the DR. Both sets of data coincide very well at
zero field. With increasing the magnetic field the specific heat
jump due to the superconducting transition moves quickly to lower
temperatures leaving a background which is consistent with that
above $T_c$ at zero field. This provides a reliable way to extract
the normal state specific heat as shown by the thick solid line
since the normal state can be described by $C/T=\gamma_n +\beta
T^2$, where the first and the second terms correspond to the normal
state electronic and phonon contribution, respectively. From the
data it is found that $\beta$ = 2.03 $mJ/ mol K^4$ and $\gamma_n$ =
41.7 $mJ/mol K^2$. In low temperature region ($\sim$ 0.2 K) the
superconducting state exhibits, however a residual value $\gamma_0
\approx$ 22.1 $mJ/ mol K^2$ indicating a non-superconducting
fraction of about 53\%. This high value of non-superconducting
fraction is however difficult to be regarded as due to an impurity
phase with completely different structure as $MgIrB$ since the XRD
data is quite clean. We thus suggest that the superconductivity
depends sensitively on the relative compositions among the three
elements and some regions without superconductivity have the
chemical composition and even the structure close to the
superconducting phase. This will be justified in the following
analysis. In any case, it is safe to conclude that the normal state
Sommerfeld constant ranges from about 19.6 to 41.7 $mJ/mol K^2$.
Further analysis suggests that the real $\gamma_n$ is close to the
upper bound of the experimental values, i.e., $\gamma_n \approx$ 41.
7 $mJ/mol K^2$.

\begin{figure}
\includegraphics[width=8cm]{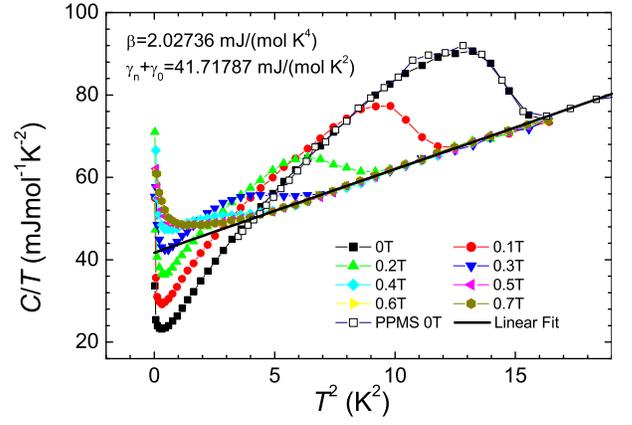}
\caption{(color online) Raw data of specific heat plotted as $C/T$
vs. $T^2$. All filled symbols represent the data taken with the DR
based on the PPMS at various magnetic fields. The open squares show
the data taken with the PPMS at zero field. The thick solid line
represents the normal state specific heat which contains both the
phonon $\gamma_{ph}$ and the electronic contributions.} \label{fig2}
\end{figure}

Next we can have an estimation on $\gamma_n$. In $MgIrB$, the
electronic conduction is dominated by the 5d band electrons of $Ir$
atoms. The DOS at $E_F$ given by the LDA band structure
calculation\cite{Wiendlocha} is about $N(E_F) = 5.51/eV spin$.
Assuming a electron-phonon coupling constant $\lambda_{e-ph}$ in the
system, one has

\begin{equation}
\gamma_n=\frac{2\pi^2}{3}N(E_F)k_B^2(1+\lambda_{e-ph}).\label{eq:1}
\end{equation}

Using the band structure value of $N(F_F)$, we have $\gamma_n$ =
25.98 $(1+\lambda_{e-ph})$$(mJ/mol K^2)$. Taking the lower bound of
the experimental value $\gamma_n$ = 19.6 $mJ/molK^2$ implies an
un-physical value $\lambda_{e-ph}$ = - 0.25. Taking however, the
upper bound of the experimental value $\gamma_n$ = 41.7, we get
$\lambda$ = 0.6. Therefore it seems that the real value of
$\gamma_n$ is close to the upper bound of the experimental values.
We can also use an alternative way to estimate $\gamma_n$. In the
dirty limit, for a type-II superconductor, one
has\cite{JaffePRB1989}

\begin{equation}
-\frac{\partial \mu_0H_{c2}}{\partial T}|_{T_c}=A \rho_n\gamma_n
\eta,\label{eq:2}
\end{equation}

where $A=3.81e/\pi^2k_B$=0.0081$(T/K)(m\Omega cm)^{-1}(mol K^2/mJ)$
for $MgIrB$. Using $- d \mu_0H_{c2}(T)/dT|_{T_c} \approx 0.3 T/ K$,
$\rho_n$ = 0.816 $m\Omega cm$, and taking $\eta =1$ for the weak
coupling case, we have $\gamma_n$ = 45 $mJ/mol K^2$ which is also
close to the upper bound of the experimental value.

In the raw data shown in Fig.2, one can see an upturn of
$\gamma=C/T$ in the very low temperature region. This upturn is
known as the Schottky anomaly, induced by lifting the degeneracy of
the states of the paramagnetic spins. We tried a two level (S=1/2)
model to fit the low temperature data but found a poor fitting
together with an extremely large Land\'{e} factor $g$ in the Zeeman
energy $g\mu_BH_{eff}$, where $\mu_B$ is the Bohr magneton,
$H_{eff}$ = $\sqrt{H^2+H_0^2}$ is the effective magnetic field which
evolves into $H_{eff}$ = $H_0$, the crystal field at zero external
field. In $MgIrB$ the most possible paramagnetic centers are from
$Ir^{4+}$ (S=5/2) or $Ir^{3+}$ (S=2). The system energy due to
Zeeman splitting in a magnetic field is\cite{Schottky}

\begin{equation}
E_{Sch}=\sum E_iexp(-E_i/k_BT)/\sum exp(-E_i/k_BT),\label{eq:3}
\end{equation}

where $E_i=M_Jg\mu_BH_{eff}$ and $M_J$ = -S, -S+1, ..., S-1, S. The
specific heat due to the Schottky effect is thus
$C_{Sch}=(n/k_B)dE_{Sch}/dT$, where $n$ represents the concentration
of the paramagnetic centers. For $S=5/2$ and $S=2$ the calculated
results are very close to each other, therefore we show only the fit
with $S=5/2$ correspoding to $Ir^{4+}$ (six levels). This method
allows us to deal with the data at zero and finite fields
simultaneously. It is known that the Schottky term should be zero at
T = 0 K. In the superconducting state, the total specific heat can
be written as: $C_{tot}=C_{nons}+C_e+C_{ph}+C_{Sch}$ with
$C_{nons}=\gamma_0T$ the contribution of the non-superconducting
regions, $C_e$ is the electronic part. In the zero temperature limit
only the contribution of the non-superconducting part is left.
Applying a magnetic field gives rise to a finite value $\Delta
\gamma_e$ to $C_e=\gamma_eT$ due to the presence of vortices.
Practically, in order to fit the Schottky term, we first remove the
phonon contribution $C_{ph}=\beta T^3$, then vertically move the
experimental data downward with a magnitude $\gamma_0$ = 22.1 $mJ/
mol K^2$ and a field induced vortex term $\Delta \gamma_e(H)$ as
shown in Table-I. Four sets of data after this treatment and the
corresponding fits to the Schottky term at the fields of $\mu_0H$ =
0.0, 0.1, 0.2, 0.6 T are shown in Fig.3. It is clear that low
temperature upturn can be well described by the Schottky effect. The
results yielded by the fitting are summarized in Table-I. One can
see that when field is beyond 0.5 T which is close to the upper
critical field, we take the total normal state value $\gamma_n$ as
the removed background which leads to a perfect fitting to the
Schottky term as shown in Fig.3(d).

\begin{table}
\caption{Fitting parameters of Schottky anomaly.}
\begin{tabular}{ccccc}
\hline \hline
$\mu_0H$(T) & $\Delta\gamma_{e}$(mJ$/$mol$\,$K$^{2}$)   &  $\mu_0H_{0}$(T)   & $n$(mJ$/$mol$\,$K)  & $g$ \\
\hline
0.0      & $0$           & $0.04$      & $6.82$        & $2$  \\
0.1      & $3.94$        & $0.04$      & $3.91$        & $2$  \\
0.2      & $7.85$        & $0.04$      & $4.82$        & $2$  \\
0.3      & $11.86$       & $0.04$      & $5.82$        & $2$  \\
0.4      & $14.80$       & $0.04$      & $6.82$        & $2$  \\
0.5      & $19.58$       & $0.04$      & $7.07$        & $2$  \\
0.6      & $19.58$       & $0.04$      & $7.07$        & $2$  \\
0.7      & $19.58$       & $0.04$      & $7.07$        & $2$  \\
 \hline \hline
\end{tabular}
\label{tab:table1}
\end{table}

After successfully removing the Schottky term and the contribution
from the non-superconducting region, we get the pure contribution
from the superconducting regions (as shown in Fig.4). Note that here
we used the value 41.7 $mJ/ mol K^2$ as the normal state Sommerfeld
constant $\gamma_n$ if it would contain only the pure
superconducting phase. One can see that the low temperature part is
flattened out below about 0.8 K when the field is zero. Actually
this flattening is already visible in the data shown in Fig.3(a)
before the Schottky term is removed. Furthermore it can also be
justified by the requirement of entropy conservation. Since the
Schottky term gives only very small contribution in the high
temperature region (above 1.5 K here), if $\gamma_e$ had a power law
instead of a flat temperature dependence here, the entropy would be
clearly not conserved yielding a large negative entropy. This is of
course unreasonable. In Fig.4 we present together the theoretical
curve for $\gamma_e$ calculated using the weak coupling BCS formula

\begin{figure}
\includegraphics[width=8cm]{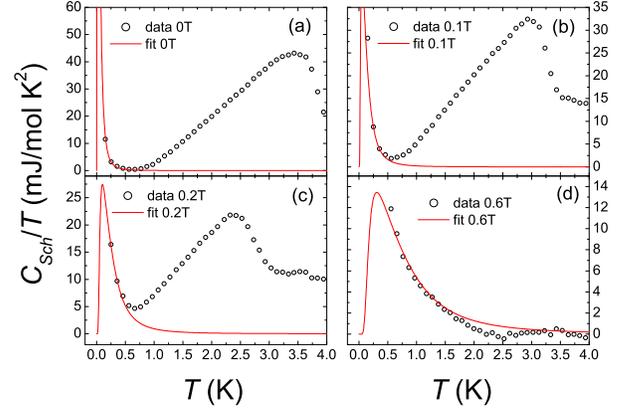}
\caption {(color online) The calculated Schottky anomaly (red solid
line) compared with the electronic specific heat (symbols) in the
low temperature regime, where the phonon term $C_{ph}/T$, field
induced term $\Delta \gamma_e$ and the nonsuperconducting term
$C_{nons}/T$ have been removed. The detailed values used for the
Schottky anomaly and the electronic contributions are given in
Table-I. } \label{fig3}
\end{figure}

\begin{figure}
\includegraphics[width=8cm]{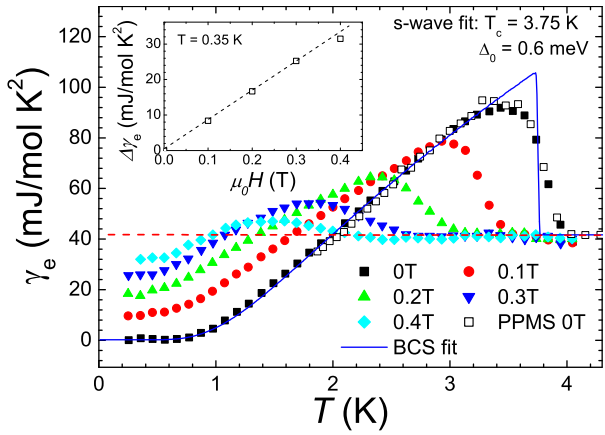}
\caption {(color online) All symbols show the temperature dependence
of the electronic specific heat in the superconducting state with
the contributions from phonon, Schottky anomaly and the
non-superconducting fraction removed. The blue solid line shows the
BCS fitting curve at zero field. The inset shows the field
dependence of the specific heat in the zero temperature limit. Note
that in this figure the normal state Sommerfeld constant $\gamma_n$
has been scaled up to $41.7 mJ/mol K^2$ as marked by the horizontal
line. } \label{fig4}
\end{figure}

\begin{equation}
\gamma_e(T) =
\frac{4N(E_F)}{k_BT^3}\int_{0}^{\hbar\omega_D}d\varepsilon[\varepsilon^2+\Delta^2(T)-\frac{T}{2}\frac{d\Delta^2(T)}{dT}]\frac{e^{\zeta/k_BT}}{(1+e^{\zeta/k_BT})^2},\label{eq:5}
\end{equation}

where $\zeta=\sqrt{\varepsilon^2+\Delta^2(T)}$. In obtaining the
theoretical fit we take the implicit relation $\Delta(T)$ derived
from the weak coupling BCS theory for an s-wave superconductor and
use the gap $\Delta_0$ and $T_c$ as two trying parameters. The
theoretical curve fits the experimental data very well leading to an
isotropic gap value $\Delta_0$ = $0.6 meV$ and $T_c$ = $3.75 K$. The
ratio $\Delta_0/k_BT_c$ = 1.86 obtained here is quite close to the
prediction for the weak coupling limit ($\Delta_0/k_BT_c$ = 1.76).
This is self-consistent with the conclusion derived from the
estimation on $\gamma_n$. In addition, the specific heat anomaly at
$T_c$ is $\Delta C_e/\gamma_nT|_{T_c} \approx $ 1.54 being very
close to the theoretical value 1.43 predicted for the case of weak
coupling. The inset in Fig.4 shows a field induced part $\Delta
\gamma_e$. In an s-wave superconductor $\Delta \gamma_e$ is mainly
contributed by the vortex cores and a linear relation $\Delta
\gamma_e \propto H \gamma_n/H_{c2}(0)$ is anticipated\cite{Hussey}
in the low field region with $\Delta_0^2/E_F \ll T \ll T_c$. This
linear relation is well demonstrated by the data below 0.4 T,
indicating another evidence of s-wave pairing symmetry.

\begin{figure}
\includegraphics[width=8cm]{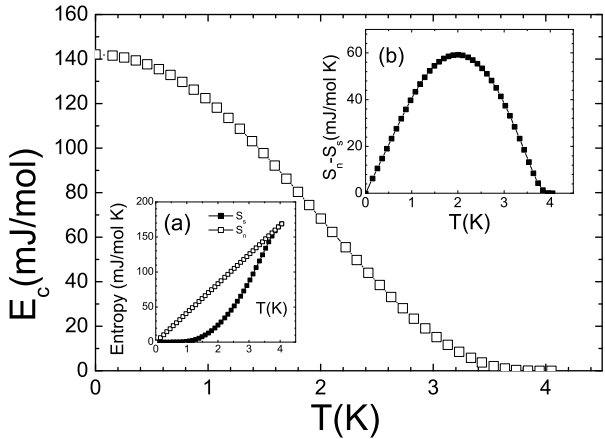}
\caption {The condensation energy calculated from the specific heat.
Inset (a) shows the entropy in the normal and superconducting state.
Plotted in inset (b) is the difference of the entropy between the
normal and superconducting state. } \label{fig5}
\end{figure}

In the following we try to estimate the superconducting condensation
energy $E_{c}$. In calculating $E_{c}$ we get the entropy difference
between the normal state and the superconducting state by
$S_n-S_s=\int_0^T(\gamma_n-\gamma_e)d T'$, then $E_{c}$ is
calculated through $E_{c}=\int_T^{4K}(S_n-S_s)dT'$. The data of
$S_n$ and $S_s$ as well as the difference between them are shown in
inset (a) and (b) of Fig.5, respectively. The main frame of Fig.5
shows the temperature dependence of the condensation energy $E_c$
which is about 142 $mJ/mol$. This value can actually be assessed by
the following equation

\begin{equation}
E_{c}=\alpha N(E_F)\Delta_0^2/2=\alpha
\frac{3}{4\pi^2}\frac{1}{k_B^2}\gamma_n\Delta_0^2\label{eq:4}
\end{equation}

For a BCS s-wave superconductor, $\alpha$ = 1. Taking $\gamma_n =
41.7 mJ/mol K^2$ and $\Delta_0$ = 0.6 meV, we found a value of
$E_{c} \approx $ 154 $mJ/mol$ which is very close to the
experimental value 142 $mJ/mol$.

Now we get down to the electron-phonon coupling. From the normal
state value we have derived that $\beta \approx$ 2.03 $mJ/mol K^4$.
Using the relation $\Theta_D$ =$(12\pi^4k_BN_AZ/5\beta)^{1/3}$,
where $N_A$ = 6.02$\times 10^{23}$ the Avogadro constant, Z=45 the
number of atoms in one unit cell, we get the Debye temperature
$\Theta_D (MgIrB)$ = 350.5 K. It is known that the Debye temperature
for metallic $Ir$ is about 36 meV (420 K)\cite{Wiendlocha}. While
the crystalline boron has a very high Debye temperature $\Theta_B$ =
100 meV (1200 K)\cite{Wiendlocha}. The rather low Debye temperature
found in our experiment $\Theta_D (MgIrB)$ = 350.5 K is close to
that of $Ir$ metal, this may suggest that the conduction electrons
from the 5d band of $Ir$ couple most strongly with the vibrations of
$Ir$ themselves. However, as we stressed before, it seems that the
$T_c$ can be tuned to higher values by changing the relative
compositions among the three elements $Mg$, $Ir$ and $B$. This may
enhance the electron-phonon coupling and/or the quasiparticle DOS at
$E_F$. The basic parameters and properties derived in this work
provide a playground for the future study in this interesting
system.

In summary, analysis on the low-temperature data in $MgIrB$ finds a
s-wave pairing symmetry with a gap in the weak coupling regime. The
conduction electrons interact primarily with the vibrations of the
$Ir$ atoms leading to a weak coupling strength $\lambda \approx$
 0.6.

\begin{acknowledgments}
% put your acknowledgments here.
We acknowledge the fruitful discussions with Tao Xiang and Junren
Shi at IOP, CAS, and Guoqing Zheng at Okayama University, Japan.
This work is supported by the National Science Foundation of China,
the Ministry of Science and Technology of China (973 project No:
2006CB601000, 2006CB921802), and Chinese Academy of Sciences
(Project ITSNEM).
\end{acknowledgments}

\end{document}